**Edge-channel interferometer at the graphene quantum Hall pn junction**

Sei Morikawa[1], Satoru Masubuchi[1,2,a], Rai Moriya[1], Kenji Watanabe[3], Takashi Taniguchi[3], and Tomoki Machida[1,2,b]

[1]*Institute of Industrial Science, University of Tokyo, 4-6-1 Komaba, Meguro, Tokyo 153-8505, Japan*

[2]*Institute for Nano Quantum Information Electronics, University of Tokyo, 4-6-1 Komaba, Meguro, Tokyo 153-8505, Japan*

[3]*National Institute for Materials Science, 1-1 Namiki, Tsukuba 305-0044, Japan*

**Abstract:** We demonstrate a quantum Hall edge-channel interferometer in a high-quality graphene pn junction under a high magnetic field. The co-propagating p and n quantum Hall edge channels traveling along the pn interface functions as a built-in Aharanov-Bohm-type interferometer, the interferences in which are sensitive to both the external magnetic field and the carrier concentration. The trajectories of peak and dip in the observed resistance oscillation are well reproduced by our numerical calculation that assumes magnetic flux quantization in the area enclosed by the co-propagating edge channels. Coherent nature of the co-propagating edge channels are confirmed by the checkerboard-like pattern in the dc-bias and magnetic-field dependences of the resistance oscillations.

[a] E-mail: msatoru@iis.u-tokyo.ac.jp
[b] E-mail: tmachida@iis.u-tokyo.ac.jp

Recent advancements in the transfer technique of atomic layers has enabled the fabrication of graphene/hexagonal boron nitride (h-BN) van der Waals heterostructures [1]. Graphene/h-BN demonstrated a large enhancement in carrier mobility due to significant suppression in charge-carrier scattering induced by extrinsic scattering sources originating from its substrate [2,3], and thus the mean free path of its carriers is in a macroscopic length scale. Consequently, ballistic carrier-transport phenomena such as electron focusing [4], negative bend resistance [5], and magnetic commensurability effect [6] have been observed in graphene/h-BN. Further, h-BN can also be used as dielectrics for top and bottom gate electrodes to control the local carrier density. An ultra-clean lateral graphene npn junction in a dual-gate h-BN/graphene/h-BN configuration has been fabricated [7–9]. As the charge carriers are transmitted with its phase memory maintained over the junction size, such devices exhibit unique coherent-transport phenomena of Dirac fermions such as Fabry–Perot interference and Klein tunneling [7,8,10].

In the presence of a strong perpendicular magnetic field, under which the electronic states in graphene are quantized into Landau levels, graphene exhibits the half-integer quantum Hall effect [11,12], and a one-dimensional conduction channel called the quantum Hall edge channel is formed along the boundary of the quantum Hall conductor. The gapless nature of the band structure in graphene allows us to form quantum Hall pn junctions [13,14]. Since the quantum Hall edge channels have opposite chirality in p and n regions, the charge carriers in the counter-circulating p and n quantum Hall edge channels travel in the same direction along the pn interface [Fig. 1(a)]. Recent experiments showed that the mixing of the co-propagating edge channels is strongly suppressed in high-mobility graphene pn junctions and that the co-propagating edge channels adiabatically transmit carriers along the pn interface [15]. On the other hand, a theory predicted the enhancement of the extrinsic edge scattering and the intrinsic inter-valley scattering near the edge [16]. Consequently, two co-propagating quantum Hall edge channels are expected to form a closed path [Fig. 1(b)]. When the coherent electronic waves are transmitted through the quantum Hall pn junction, we expect that the interference will depend on the magnetic flux penetrating the closed path; thus the transport coefficient will be modulated by changing the external magnetic field or the area enclosed by the co-propagating edge channels.

In this Letter, we study an oscillatory magnetoresistance in a graphene quantum Hall npn junction by employing a dual-gate graphene encapsulated by h-BN. The trajectories of peak and dip of the resistance oscillation are well reproduced by our numerical calculation that assumes magnetic flux quantization in the area enclosed by the co-propagating quantum Hall edge channels. Furthermore, the dependence of conductance on the source–drain bias voltage and magnetic field exhibits checkerboard-like patterns, confirming coherent interference, which is a prerequisite of the novel resistance oscillations under study. These results suggest that a naturally formed edge-channel interferometer exists at graphene quantum Hall pn junction as exemplified in Fig. 1(b).

Our dual-gated graphene device was fabricated as follows. First, a 50-nm-thick h-BN flake was deposited on a Si wafer (with a 290-nm-thick thermally oxidized layer of $SiO_2$) through the mechanical exfoliation of h-BN. Then, monolayer graphene exfoliated from Kish graphite was transferred onto the h-BN flake by using the mechanical transfer technique [2,17]. In order to remove surface contaminations, the sample was annealed in Ar/H$_2$ (97:3) atmosphere at 350°C. Next, another 50-nm-thick h-BN flake was deposited on top. Standard e-beam lithography, e-beam evaporation, and a subsequent lift-off process were used to fabricate the Pd (70 nm) metal contacts and the top gate electrode. Finally, the sample was annealed in an Ar/H$_2$ (97:3) atmosphere at 300°C. The width and length of the top-gated region were 1700 and 430 nm, respectively. The two-terminal resistance $R$ was measured using a standard lock-in technique with an alternating current $I_\text{ac}$ = 5 nA at a frequency of 18 Hz in a $^4$He-cooled variable-temperature insert at $T$ = 1.5 K. A magnetic field $B$ of up to 9 T was applied perpendicularly to the graphene by using a superconducting magnet.

A schematic of our device is shown in Fig. 1(a). A more detailed device structure is shown in Fig. S1(a) in the supplementary information [18]. The charge-carrier densities in the top-gated and uncovered regions $n_\text{t}$ and $n_\text{b}$ were tuned by applying back-gate and top-gate bias voltages $V_\text{b}$ and $V_\text{t}$ [19–22]. The two-dimensional plot of $R$ as a function of $V_\text{b}$ and $V_\text{t}$ (Fig. S1(b) in the supplementary information [18]) reveals a small residual doping of $\Delta n \sim 10^{11}$ cm$^{-2}$ and a high charge-carrier mobility of 71000 cm$^2$/Vs at $T$ = 1.5 K. The value of top-gate and back-gate capacitance extracted from magnetotransport measurements were in good agreement with geometrical considerations. The device exhibited Fabry–Perot-type resistance oscillations and Klein tunneling in an npn cavity configuration (Figs. S1(c) and S1(d) in the supplementary information [18]). These observations indicate that coherent transport is achieved in the top-gated graphene region with the phase coherence length well exceeding the cavity length of 430 nm. In the following, we reveal that this coherent npn junction functions as a built-in Aharanov-Bohm interferometer in high magnetic fields.

The two-terminal resistance of the dual-gated graphene npn junctions in quantum Hall regime is determined by two different quantities. One is the Landau-level filling factors $\nu_t$ and $\nu_b$ in p and n regions. The other is the inter-edge-channel scattering rate between the p and n co-propagating quantum Hall edge channels. In a conventional graphene npn junction on a SiO$_2$/Si substrate [14], the p and n edge channels are fully mixed at the pn interface because of the high inter-edge-channel scattering rate, and the two-terminal resistance is a classical addition of the three quantized Hall conductors in series as

$$R = \frac{h}{e^2}\left(\frac{1}{|\nu_t|} + \frac{2}{|\nu_b|}\right). \quad (1)$$

On the other hand, in a high-mobility clean graphene npn junction on h-BN, the mixing of co-propagating edge channels is strongly suppressed, and carriers are adiabatically transmitted through the edge channels. This is due to the existence of a $\nu = 0$ insulating area between the p and n quantum Hall edge channels; the valley/spin degeneracy of the zero-energy Landau level is lifted, and the $\nu = 0$ insulating state is stabilized in high-mobility graphene under a high magnetic field. Such a region with $\nu = 0$ separates the co-propagating edge channels and reduces overlap between their wave functions [15]. Therefore, it is expected that the two-terminal resistance in a high-mobility clean graphene npn junction on h-BN no longer follows Eq. (1) and is larger than the value given by Eq. (1).

Fig. 2(a) presents magnetotransport measurements in the npn regime, where the two-terminal resistance $R$ is plotted for the charge carrier densities in the gated regions of $n_t =$ 0.5, 0.7, 0.9, 1.1, 1.3, 1.5, and $1.7 \times 10^{12}$ cm$^{-2}$ (bottom to top) with $n_b = -2.2 \times 10^{12}$ cm$^{-2}$. For $n_t = 0.5 \times 10^{12}$ cm$^{-2}$ and $B > 2$ T, $R$ became larger than the value expected for fully mixed quantum Hall edge channels (Eq. (1)), which suggests that the co-propagating quantum Hall edge channels here are decoupled and travel adiabatically along the pn interface [15]. Moreover, $R$ started to oscillate as a function of $B$. On increasing $B$ further, the oscillation amplitude of $R$ gradually decreases, and oscillation stops for $B > 4$ T. When $n_t$ is increased, the $R$ curves qualitatively show the same behavior, but the magnetic-field range in which the resistance shows oscillatory behavior is shifted to higher magnetic fields.

Figs. 2(b) and 2(c) show the oscillation period $\Delta B$ of $R$, derived from Fig. 1(a), as a function of $B$ for $n_t = 1.3 \times 10^{12}$ cm$^{-2}$ and as a function of $n_t$ for $B = 4$ T, respectively. In these plots, $\Delta B$ is plotted in the ranges in which the oscillation amplitude is sufficiently large, i.e., for $\Delta R_{osc} > 6$ k$\Omega$. It can be seen that $\Delta B$ decreases with $B$ and increases with $n_t$. The magnetoresistance oscillations are reminiscent of Aharanov–Bohm oscillation. However, if $\Delta B$ is related to the area $S$ enclosed by the interfering path according to the relation $\Delta BS = \phi_0$ [23], $\Delta B \sim 0.4$ T in our experiment corresponds to $S \sim 10^4$ nm$^2$, which is smaller than the area of the top-gated region by two or three orders of magnitude. Further, to account for the dependence of $\Delta B$ on $B$ and $n_t$ in Figs. 2(b) and 2(c), $S$ must increase with $B$ and decrease with $n_t$. The top-gated region cannot satisfy these conditions.

In the following, we show that the area $S$ surrounded by co-propagating quantum Hall edge channels in the vicinity of the pn junction describes our experimental findings. The local charge-carrier density $n(x)$ across the pn interface gradually changes because of the finite thickness of the h-BN layers (Figs. 1(c) and 1(d)). Under a high magnetic field, the quantum Hall conductor is separated into conductive regions and localized regions, which correspond to the quantum Hall edge channels and quantum Hall insulating regions, respectively. Especially, the electronic states around $n(x) = 0$ become insulating (yellow areas in Figs. 1(c) and 1(d)) because of the breaking of Landau-level degeneracy, the condition for which can be expressed as $|n(x)| < \nu_c B/\phi_0$, where $\nu_c$ is the critical filling factor [24–27]. Thus, the area $S$ of the insulating region can be derived from $B$, $\nu_c$, and the profile of $n(x)$.

When $B$ is increased, the range of $n(x)$ which satisfies $|n(x)| < \nu_c B/\phi_0$ also increases. Consequently, the area $S$ surrounded by quantum Hall edge channels becomes larger (Fig. 1(c)). On the other hand, increasing the carrier densities in the p and n regions increases the slope of $n(x)$. Consequently, $S$ is decreased so that the relationship $|n(x)| < \nu_c B/\phi_0$ is maintained (Fig. 1(d)). Considering the relation $\Delta B S = \phi_0$, these changes in $S$ with respect to $B$ and $n$ are qualitatively consistent with the change in $\Delta B$ shown in Figs. 2(b) and 2(c).

We quantitatively derived $S$ through a numerical simulation using the finite-element method. The charge-carrier density profile $n(x)$ was calculated by considering the realistic geometry of our device. The calculated results for various experimental conditions are shown in Figs. S2(a)–S2(c) in the supplementary information [18]. From the top-gated region ($x > 0$) to the uncovered region ($x < 0$), $n(x)$ gradually changes within a length scale of 100 nm, and we derived the value of $S$ as shown in Figs. S2(d)–S2(f) in the supplementary information [18]. It can be seen that $S$ is on the order of $10^4$ nm$^2$, which is consistent with $\Delta B \sim 0.4$ T in our experiments. These calculations involve only one adjustable parameter, $\nu_c$; we set its value as 0.4, which is close to the values reported in experiments [24–27].

The right panel of Fig. 3(a) shows the locations where the magnetic flux penetrating the insulating region is integer multiples of the magnetic-flux quantum $\phi_0$ as $\Phi(B, n_t) = BS = N\phi_0$ ($N = 1, 2, \ldots, 20$) in the $B$–$n_t$ plane; here we used $S$ in Fig. S2(d). These curves were compared with the experimental data in the left panel of Fig. 3(a), which shows a color plot of $R$ as a function of $B$ and $n_t$ for the npn regime. The oscillatory behavior of $R$ in the magnetic-field range of 2 T $< B <$ 8 T corresponds to the resistance oscillations shown in Fig. 2(a). In this range, the positions of peak and dip structures were well reproduced by the calculated curves in Fig. 3(a). In addition, the observed oscillatory $R$ obtained as a function of different parameters ($n_b$, $B$) and ($V_b$, $B$), which correspond to the cases of Figs. S2(e) and S2(f), respectively, were also explained by the calculated curves (Figs. 3(b) and 3(c)) using the same value of $v_c = 0.4$. Therefore, the oscillatory behavior of $R$ can be fully explained by considering magnetic flux that penetrates the insulating area between the co-propagating p and n quantum Hall edge channels; these results show the presence of the Aharanov–Bohm interferometer consisting of co-propagating edge channels at the graphene quantum-Hall pn junction as shown in Fig. 1(b).

The application of a dc bias to the device changes the chemical potential of the edge channel, thereby changing its wave vector as well. Therefore, if quantum interference exists between the quantum Hall edge channels, the interference condition oscillates with the dc-bias voltage $V_{dc}$, as has been observed in a two-dimensional electron system in a modulation-doped AlGaAs/GaAs heterostructure [28]. Fig. 4 shows the color plot of the differential conductance $G = dI/dV_{dc}$ as a function of $V_{dc}$ and $B$ measured by superposing a small ac bias voltage of $V_{ac} = 100$ μV on $V_{dc}$. The differential conductance oscillates as a function of $V_{dc}$, and the color plot of $dI/dV_{dc}$ shows a clear checkerboard-like pattern.

This result constitutes further evidence indicating that the observed oscillatory behavior of magnetoresistance is attributable to the quantum interference between co-propagating p and n quantum Hall edge channels.

**Acknowledgments**

The authors acknowledge M. Onuki, M. Arai, T. Yamaguchi, M. Koshino, Y. Hamamoto, and Y. Hatsugai for technical support and discussions. This work was partly supported by the Grant-in-Aid for Scientific Research on Innovative Areas "Science of Atomic Layers" of the Ministry of Education, Culture, Sports, Science and Technology (MEXT); the Project for Developing Innovation Systems of MEXT; the Grants-in-Aid for Scientific Research from the Japan Society for the Promotion of Science (JSPS). S. Morikawa acknowledges the JSPS Research Fellowship for Young Scientists.

**Figure captions**

Fig. 1.

(Color online) (a) Schematic of our device and edge-channel configuration in a quantum Hall state. (b) Expected built-in interfering path between co-propagating edge channels at the pn junction. The edge channels are assumed to be predominantly mixed at the edge of the interface. (c) and (d) Qualitative explanation of the $B$ and $n$ dependences of the area of the $\nu = 0$ insulating region $S$ (yellow region), where $|n(x)| < v_c B/\phi_0$ is satisfied.

Fig. 2

(a) Two-terminal resistance $R$ as a function of $B$ for $n_t$ = 0.5, 0.7, 0.9, 1.1, 1.3, 1.5, and $1.7 \times 10^{12}$ cm$^{-2}$ (bottom to top). (b) and (c) The oscillation period $\Delta B$ as a function of $B$ for $n_t = 1.3 \times 10^{12}$ cm$^{-2}$ and as a function of $n_t$ for $B = 4$ T.

Fig. 3

(Color online) [Left panels] Color plots of the two-terminal resistance $R$ as a function of (a) ($n_t$, $B$) for $n_b = -2.2 \times 10^{12}$ cm$^{-2}$, (b) ($n_b$, $B$) for $n_t = 1.3 \times 10^{12}$ cm$^{-2}$, and (c) ($V_b$, $B$) for $V_t = -8$ V. [Right panels] Calculated locations where the magnetic flux penetrating the insulating region is integer multiples of the magnetic flux quantum.

Fig. 4

(Color online) Color plot of differential conductance $G = dI/dV_{dc}$ as a function of dc bias voltage $V_{dc}$ and magnetic field $B$ for $(n_t, n_b) = (2.5 \times 10^{12}, -1.1 \times 10^{12})$ cm$^{-2}$.

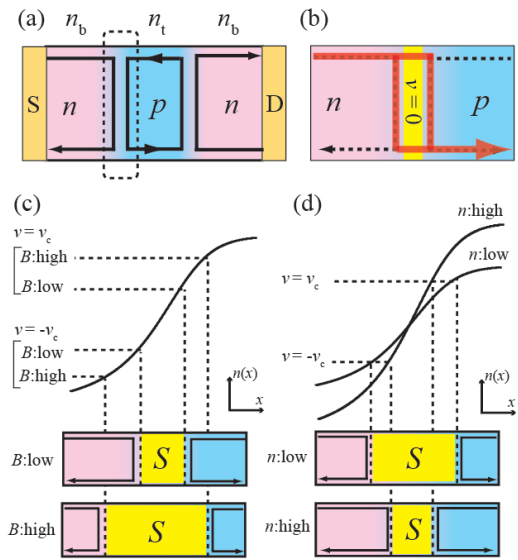

Figure 1

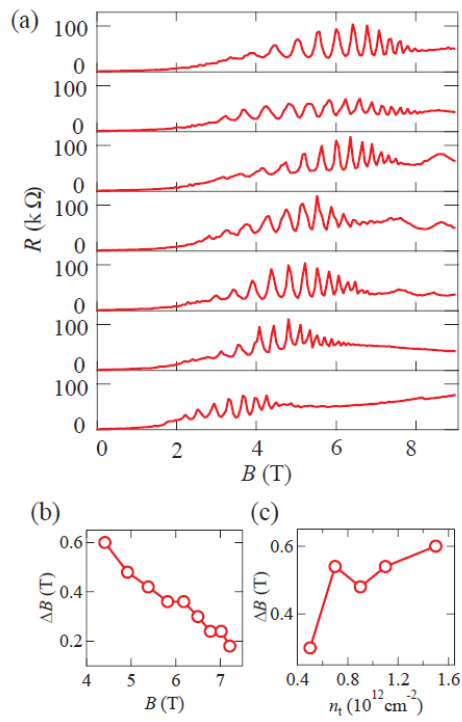

Figure 2

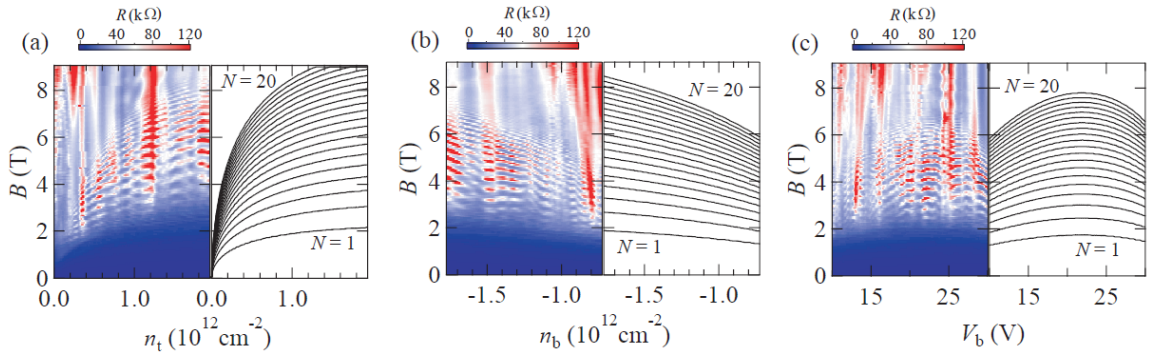

Figure 3

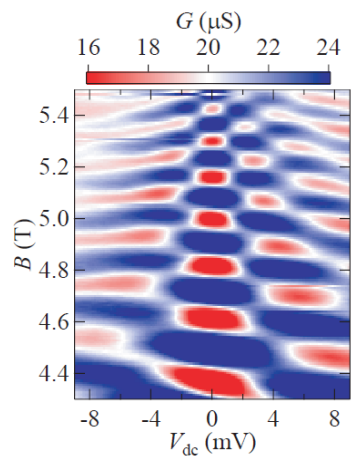

Figure 4

**Edge-channel interferometer at the graphene quantum Hall pn junction**

Sei Morikawa[1], Satoru Masubuchi[1,2], Rai Moriya[1], Kenji Watanabe[3], Takashi Taniguchi[3], and Tomoki Machida[1,2]

[1]*Institute of Industrial Science, University of Tokyo, 4-6-1 Komaba, Meguro, Tokyo 153-8505, Japan*

[2]*Institute for Nano Quantum Information Electronics, University of Tokyo, 4-6-1 Komaba, Meguro, Tokyo 153-8505, Japan*

[3]*National Institute for Materials Science, 1-1 Namiki, Tsukuba 305-0044, Japan*

**Carrier transport properties in a dual-gated graphene pnp junction**

A schematic of the dual-gated graphene pnp junction in the present work is shown in Fig. S1(a). The charge-carrier densities in the top-gated and uncovered regions $n_t$ and $n_b$ were tuned by applying back-gate and top-gate bias voltages $V_b$ and $V_t$. The two-dimensional plot of two-terminal resistance as a function of $V_b$ and $V_t$ is shown in Fig. S1(b). The device exhibited Fabry–Perot-type resistance oscillations (Figs. S1(c)) and Klein tunneling (Figs. S1(d)) in the npn cavity configuration. These observations indicate that coherent transport is achieved in the top-gated graphene region.

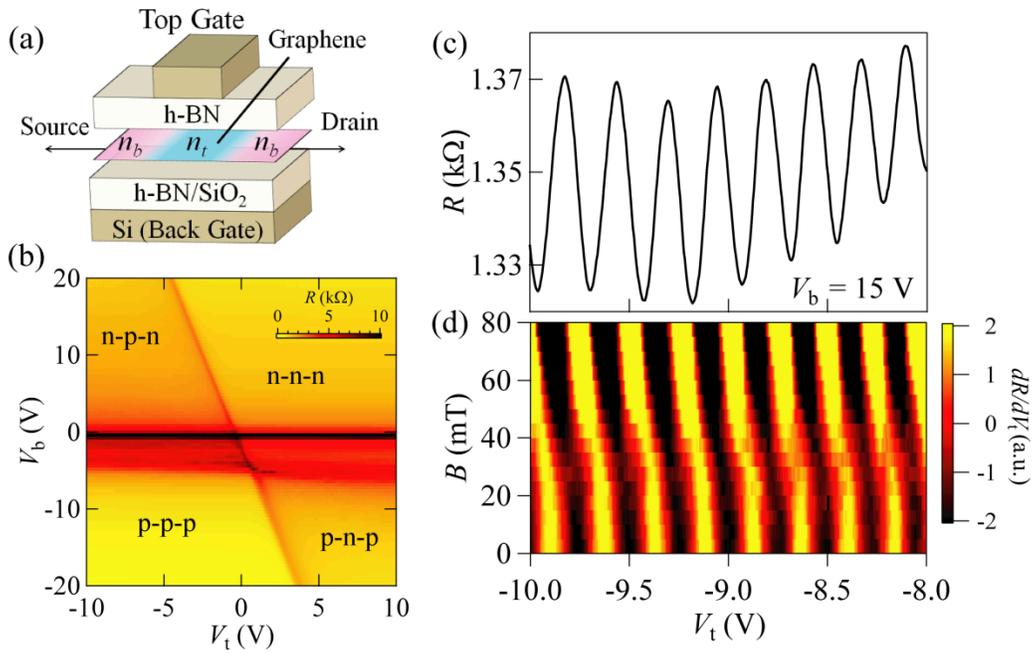

Fig. S1

(Color online) (a) Schematic of our device. (b) Two-terminal resistance $R$ as a function of $V_t$ and $V_b$ measured at $T = 1.5$ K. (c) Line cut of (b) at $V_b = 15$ V. (d) Color plot of $dR/dV_t$ as a function of $V_t$ and $B$.

**Numerical calculation of the carrier-density profile in a graphene pn junction using finite-element method**

The charge-carrier density profile $n(x)$ was calculated by considering the realistic geometry of our device. The calculated results for various experimental conditions are shown in Figs. S2(a)–S2(c). The area $S$ of the insulating region between co-propagating quantum Hall edge channels is derived as shown in Figs. S2(d)–S2(f).

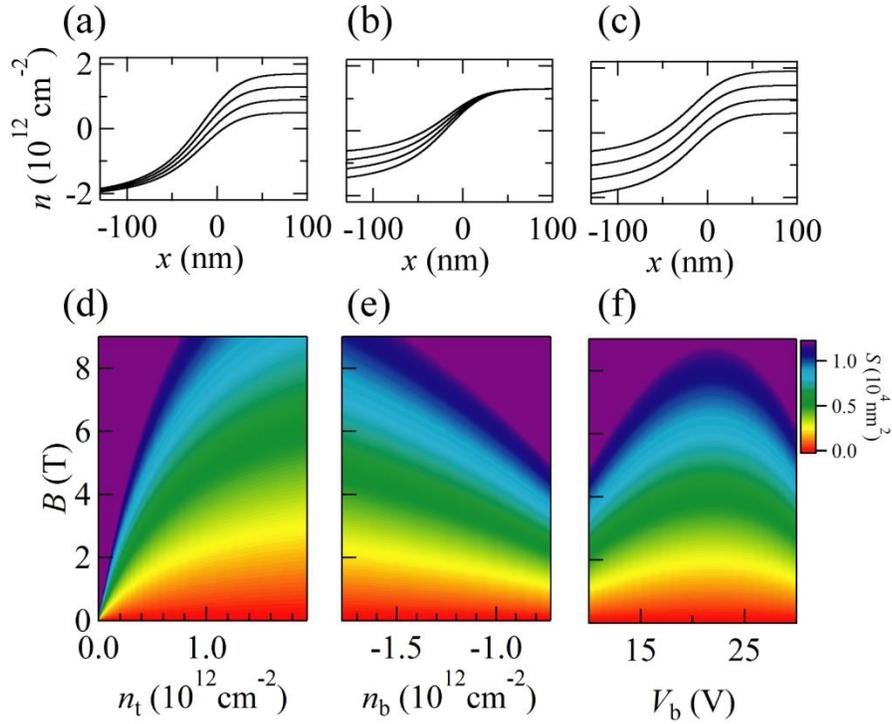

Fig. S2

(Color online) (a)–(c) Calculated carrier-density profiles $n(x)$ for (a) $n_t = 0.5 \times 10^{12}$, $0.9 \times 10^{12}$, $1.3 \times 10^{12}$, and $1.7 \times 10^{12}$ cm$^{-2}$ for $n_b = -2.2 \times 10^{12}$ cm$^{-2}$, (b) $n_b = -1.7 \times 10^{12}$, $-1.4 \times 10^{12}$, $-1.1 \times 10^{12}$, and $-0.8 \times 10^{12}$ cm$^{-2}$ for $n_t = 1.3 \times 10^{12}$, and (c) $V_b = 29, 23, 17$, and 11 V for $V_t = -8$ V (bottom to top). (d)–(f) Color plots of the calculated values of $S$ as a function of (d) $(n_t, B)$ for $n_b = -2.2 \times 10^{12}$ cm$^{-2}$, (e) $(n_b, B)$ for $n_t = 1.3 \times 10^{12}$ cm$^{-2}$, and (f) $(V_b, B)$ for $V_t = -8$ V, corresponding to the cases of (a)-(c).